\documentclass[preprint]{ptephy_v2}
\usepackage{hyperref}
\usepackage{lineno}

\begin{document}

\title{First observation of the $\gamma$-ray beam production by the backward Compton scattering of reflected synchrotron radiation in the extreme ultraviolet range}

%%%% To generate auto affiliation numbers please use \author{}\affil{} command

\author[1,2]{N.~Muramatsu \thanks{mura@impcas.ac.cn}}
\affil[1]{\small Institute of Modern Physics, Chinese Academy of Sciences, Lanzhou, 730000, China}
\affil[2]{\small Research Center for Nuclear Physics, Osaka University, Ibaraki, Osaka 567-0047, Japan}

\author[3]{M.~Miyabe}
\affil[3]{\small Research Center for Accelerator and Radioisotope Science, Tohoku University, Sendai, Miyagi 982-0826, Japan}

\author[3]{M.~Okabe}
%\affil[3]{\small Research Center for Accelerator and Radioisotope Science, Tohoku University, Sendai, Miyagi 982-0826, Japan}

\author[2]{S.~Dat\'{e}}
%\affil[2]{\small Research Center for Nuclear Physics, Osaka University, Ibaraki, Osaka 567-0047, Japan}

\author[4]{T.~Harada}
\affil[4]{\small Laboratory of Advanced Science and Technology for Industry, University of Hyogo, Kamigori, Hyogo 678-1205, Japan}

\author[4]{K.~Kanda}
%\affil[4]{\small Laboratory of Advanced Science and Technology for Industry, University of Hyogo, Kamigori, Hyogo 678-1205, Japan}

\author[5]{S.~Miyamoto}
\affil[5]{\small Institute of Laser Engineering, Osaka University, Suita, Osaka 565-0871, Japan}

\author[2]{H.~Ohkuma \thanks{Deceased.}}
%\affil[2]{\small Research Center for Nuclear Physics, Osaka University, Ibaraki, Osaka 567-0047, Japan}

\author[3]{H.~Shimizu}
%\affil[3]{\small Research Center for Accelerator and Radioisotope Science, Tohoku University, Sendai, Miyagi 982-0826, Japan}

\author[2]{S.~Suzuki}
%\affil[2]{\small Research Center for Nuclear Physics, Osaka University, Ibaraki, Osaka 567-0047, Japan}

\author[3]{A.~Tokiyasu}
%\affil[3]{\small Research Center for Accelerator and Radioisotope Science, Tohoku University, Sendai, Miyagi 982-0826, Japan}

%\author{Insert third author name here}
%\author[3]{Insert fourth author name here} %%% Use optional bracket [3] to change the respective address
%\affil{Insert third author address here}

%\author{Insert last author name here\thanks{These authors contributed equally to this work}}
%\affil{Insert last author address here}

%%% To include the collaborator name... Please use the command "\collaborator"
%%% For example: \collaborator{ATLAS Collaboration}

\begin{abstract}%
Compton scattering of photons off high-energy electrons is a fundamental quantum mechanical process widely utilized to produce a $\gamma$-ray beam for scientific research. Instead of injecting laser light into a storage ring as a conventional way, we have developed an innovative method to achieve drastically higher energies approaching the ring energy by the backward Compton scattering of extreme ultraviolet (EUV) light. In this method, $92$~$\mathrm{eV}$ photons obtained from an undulator in a storage ring were reflected back to the original ring using a Mo/Si multilayer mirror. Consequently, $\gamma$-ray beam production through the EUV light Compton scattering using reflected synchrotron radiation was observed for the first time in a demonstration experiment conducted at the $1$~$\mathrm{GeV}$ ring, NewSUBARU. The measured energy spectrum was well reproduced by a theoretical calculation with the maximum energy of $0.543$~$\mathrm{GeV}$. The production rate was $1.4 \pm 0.1$~kcps for the energies above $0.160$~$\mathrm{GeV}$. This rate was quantitatively explained by the luminosity and the scattering cross section. The present work paved the way to create a new $\gamma$-ray beam source for future applications such as hadron photoproduction experiments.
\end{abstract}

%\subjectindex{backward Compton scattering, $\gamma$-ray beam, hadron photoproduction, electron storage ring, undulator radiation, extreme ultraviolet light}

\maketitle

\section{Introduction} \label{sec1}

   The production of a $\gamma$-ray beam via backward Compton scattering from high-energy electrons was observed in 1960s using laser light (laser Compton scattering, LCS) \cite{lcsfob01, lcsfob02} as theoretically predicted in pioneer works \cite{lcspre01, lcspre02}. LCS is now in practical use at many electron storage rings by injecting infrared (IR) to ultraviolet (UV) laser light to produce a $\mathrm{MeV}$ \cite{mevgex01, mevgex02, mevgex03, mevgex04, mevgex05, mevgex06, mevgex07, mevgex08, mevgex09, mevgex10} to $\mathrm{GeV}$ \cite{gevgex01, gevgex02, gevgex03, gevgex04, gevgex05, gevgex06, gevgex07, gevgex08} $\gamma$-ray beam for physics research \cite{phyrex01, phyrex02, phyrex03, phyrex04, phyrex05, phyrex06, phyrex07, phyrex08} and industrial application \cite{iappex01, iappex02, iappex03}. This method has an advantage in significantly increasing the intensity fraction of a higher energy component, which is usually interesting for the use in the above scientific purposes, whereas the $\gamma$-ray production via bremsstrahlung provides an exponentially decreasing spectrum at higher energies. Furthermore, LCS facilities can be constructed as one of the beamlines in a storage ring, enabling to advance experimental projects in relatively low-cost manners.

\begin{figure*}[hbtp]
 \begin{tabular}{cc}
  \begin{minipage}{0.47\textwidth}
   \centering
   \includegraphics[height=6cm]{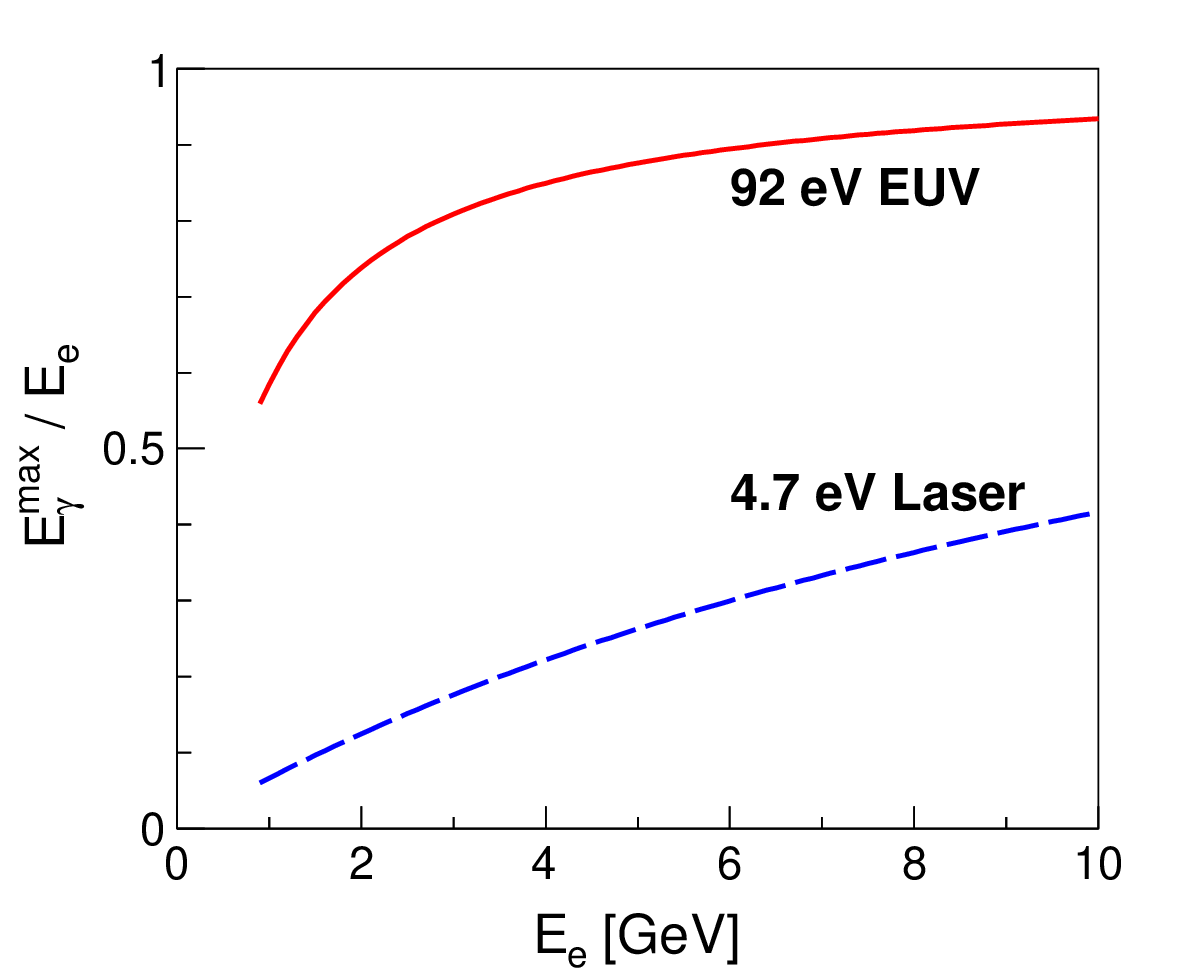}
   \caption{The ratios of the maximum $\gamma$-ray energy $E_\gamma^{max}$ to the electron beam energy $E_e$ in the backward Compton scattering processes with the incident photon energies of $4.7$ and $92$~$\mathrm{eV}$ (blue dashed and red solid lines, respectively).}
   \label{fig1}
  \end{minipage}
  \hspace{8mm}
  \begin{minipage}{0.47\textwidth}
   \centering
   \includegraphics[height=6cm]{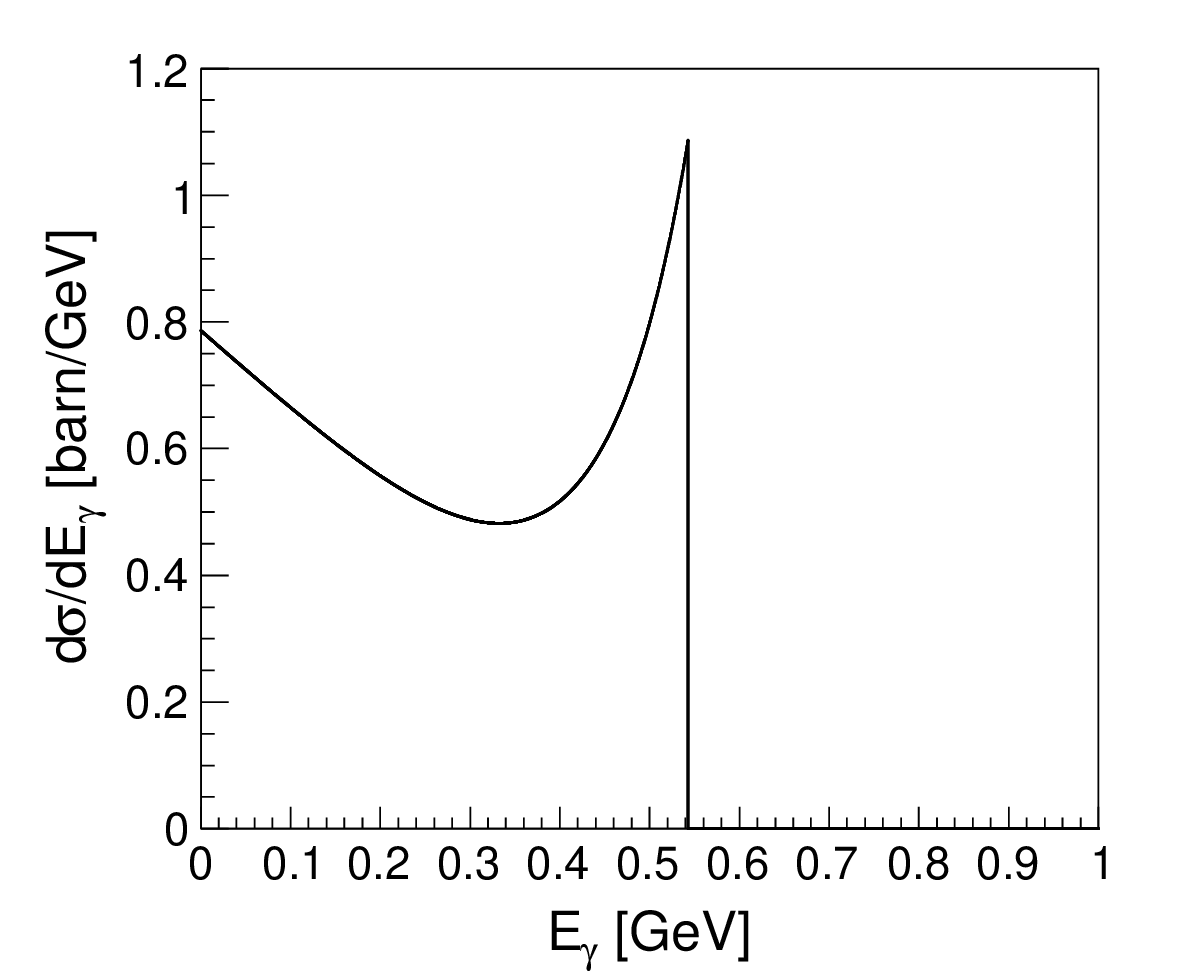}
   \caption{Backscattered $\gamma$-ray energy spectrum theoretically calculated in the form of differential cross sections for the demonstration experiment of extreme ultraviolet (EUV) light Compton scattering at NewSUBARU.}
   \label{fig2}
  \end{minipage}
 \end{tabular}
\end{figure*}

In the relativistic kinematics for the head-on collision of an incident photon and a high-energy electron possessing the energies of $k_i$ and $E_e$, respectively, the maximum energy of a scattered $\gamma$-ray $E_\gamma^{max}$ (Compton edge) is attained for the scattering angle of $0^\circ$ relative to the initial electron:
\begin{equation} \label{eq1}
  E_\gamma^{max} = \frac{4 {E_e}^2 k_i}{{m_e}^2 + 4 E_e k_i} ,
\end{equation}
where $m_e$ represents the electron mass \cite{lcskin01, lcskin02}. While $E_\gamma^{max}$ can be raised up with a large $k_i$, the availability of the highest energy laser with a high power ($\gtrsim 1$~$\mathrm{W}$) and frequency (a few tens $\mathrm{MHz}$ to CW) output is currently limited in the range up to $4.66$~$\mathrm{eV}$ (the wavelength $\lambda$ of $266$~$\mathrm{nm}$). Thus, the energies obtainable by LCS are below the blue dashed line of Fig.~\ref{fig1}, which shows that the energy transfer efficiency from an accelerated electron to a scattered photon is less than a few tens \% over a wide range of the storage ring energy $E_e$. However, further increase of $E_\gamma^{max}$ by pushing up the energy transfer efficiency is desired to promote some of research projects, particularly for hadron photoproduction experiments. For instance, the production of higher-mass particles, whose properties have not been clarified well, becomes possible if a $\gamma$-ray beam with a higher $E_\gamma^{max}$ is obtained by creating a method to inject seed photons with a larger $k_i$ into a several $\mathrm{GeV}$ storage ring. In addition, base facilities to study hadron photoproduction in various energy regions can be operated by adopting the same method at $1$--$3$~$\mathrm{GeV}$ storage rings, whose construction is the current trend in achieving an excellent electron-beam emittance. A new $\gamma$-ray beam source to replace LCS is highly called for to enable the energy extension.

So far, the injection of X-rays into a storage ring has been considered as a possible solution to increase $E_\gamma^{max}$ \cite{lcspre02, xcspre01, xcspre02}, but such $\gamma$-ray beam production has not been successful despite ambitious efforts \cite{xcspre03} because of difficulties in the generation of intense X-rays and their optical control. The method to use a free-electron laser (FEL) inside a resonator cavity, suggested in Ref.~\cite{xcspre02}, has only been realized with the incident photon energies up to $7.1$~$\mathrm{eV}$ ($\lambda = 175$~$\mathrm{nm}$) at the HI$\gamma$S facility to produce $\gamma$-ray beams in the energy range below $120$~$\mathrm{MeV}$ \cite{higs}. For a breakthrough to accomplish the drastic energy extension of a $\gamma$-ray beam, we focus on $92$~$\mathrm{eV}$ extreme ultraviolet (EUV) light ($\lambda = 13.5$~$\mathrm{nm}$), for which a Molybdenum-Silicon (Mo/Si) multilayer mirror has been developed to achieve a high reflectance at an incident angle of $0^\circ$ \cite{mlmirror}. This EUV light is available from a storage-ring undulator, and can be reflected back by the Mo/Si mirror whose refractive surface is concave to make a focus at the Compton scattering point. The electron energy in a storage ring must be efficiently transferred to a photon by the EUV light Compton scattering (EUVCS), as shown by the red solid line of Fig.~\ref{fig1}. Our method can increase $E_\gamma^{max}$ up to $80$--$90$\% for the storage ring energy of $3$~$\mathrm{GeV}$ or more. The total cross section of Compton scattering for $92$~$\mathrm{eV}$ photons on $\mathrm{GeV}$ electrons decreases only by a factor of $2$ to $4$ compared with the UV laser injection. Actually, the occurrence of a Compton scattering phenomenon using about 100 eV light has been claimed by an experiment conducted at the VEPP-2M collider in 1980s, although its purpose is limited to the polarization measurement for colliding electron and positron beams \cite{nima314-15}. Furthermore, our idea can reasonably achieve the management of all the production processes for incident EUV photons and output $\gamma$-rays at one beamline of a storage ring.

As discussed in \ref{app1}, the energy of a produced $\gamma$-ray ($E_\gamma$) becomes lower if its scattering angle is larger. (See Eq.~\ref{eq3} and Fig.~\ref{fig6}(b).) The $E_\gamma$ spectrum widely distributes from zero to the Compton edge while $\gamma$-rays are strongly boosted within a narrow cone at $\approx 0.1$~$\mathrm{GeV}$ or higher energies. Usually the energies of $\mathrm{GeV}$ $\gamma$-rays can be measured event-by-event by analyzing the momenta of recoil electrons in Compton scattering, but we have directly obtained the $E_\gamma$ spectrum by putting an electromagnetic calorimeter on the beam to confirm a $\gamma$-ray production signal in the demonstration experiment explained below. This article describes the development of a new $\gamma$-ray beam source by EUVCS (Sec.~\ref{sec2}), the observation of $\gamma$-ray beam production using the developed method (Sec.~\ref{sec3}), and the prospect of the EUVCS application in storage rings (Sec.~\ref{sec4}). A summary follows them in the end (Sec.~\ref{sec5}).

\section{Development of a new $\gamma$-ray beam source} \label{sec2}

The properties of EUV light are similar to those for soft X-rays, and the optical handling procedures for them are common in many cases. Therefore, detectors and devices for the production and confirmation of a $\gamma$-ray beam due to EUVCS were developed at the X-ray beamline BL07 \cite{nsbl07} of NewSUBARU \cite{nsring}, which is a $1$ or $1.5$~$\mathrm{GeV}$ electron storage ring. The head-on collision of $92$~$\mathrm{eV}$ photons on $350$~$\mathrm{mA}$ electrons with $E_e = 0.949$~$\mathrm{GeV}$ ($1$~$\mathrm{GeV}$ operation) should result in the $E_\gamma$ spectrum calculated by the Klein-Nishina formula \cite{landau} as shown in Fig.~\ref{fig2}, where $E_\gamma^{max}$ reaches $0.543$~$\mathrm{GeV}$. BL07 has a $2.28$~$\mathrm{m}$-long undulator for X-ray radiation with a period length of $7.6$~$\mathrm{cm}$. According to a simulation by SRW \cite{srwsim}, it radiates $92$~$\mathrm{eV}$ EUV photons with a flux of $4.11 \times 10^{14}$~$\mathrm{s}^{-1} \cdot (0.1\% \, \rm{bandwidth})^{-1}$ as the first harmonic wave when the K-value \cite{kvalue1, kvalue2} is set to $0.8$ by adjusting the strength of a magnetic field with the undulator gap setting. All the detectors and devices placed downstream of the undulator to handle the EUV light were operated under the ultra-high vacuum of $10^{-6}$~$\mathrm{Pa}$ in the same way as the X-ray case. In the experiments during the development, the radiated EUV light was directed to a downstream branch called ``A'' (BL07A) by a switching mirror with the incident angle of $87^\circ$. This mirror is made of silicon with a spherical surface coated by platinum, having a reflectance ($R$) of $88$\%. The EUV light that is reflected back by the Mo/Si multilayer mirror for the injection into the original ring travels the same path as radiated light in the opposite direction.

\begin{figure*}[hbtp]
 \centering
 \includegraphics[width=15cm]{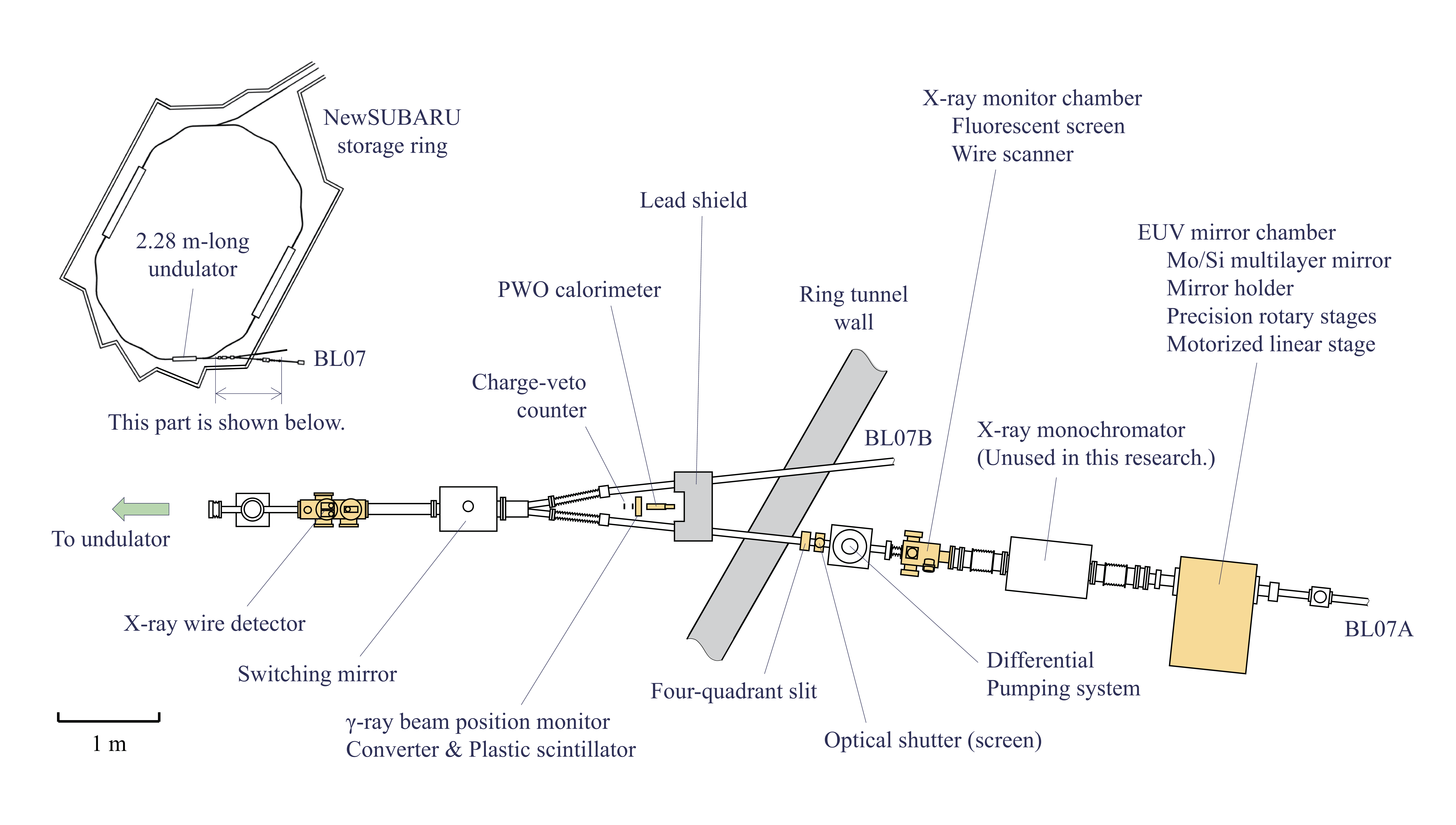}
 \caption{Experimental setup for the extreme ultraviolet (EUV) light Compton scattering. The top-left part shows a top view of the NewSUBARU ring. An expanded view of BL07A is displayed in the remaining part.}
 \label{fig3}
\end{figure*}

\begin{figure}[htbp]
 \begin{tabular}{ll}
  \begin{minipage}{0.5\linewidth}
   \centering
   \includegraphics[width=6.5cm]{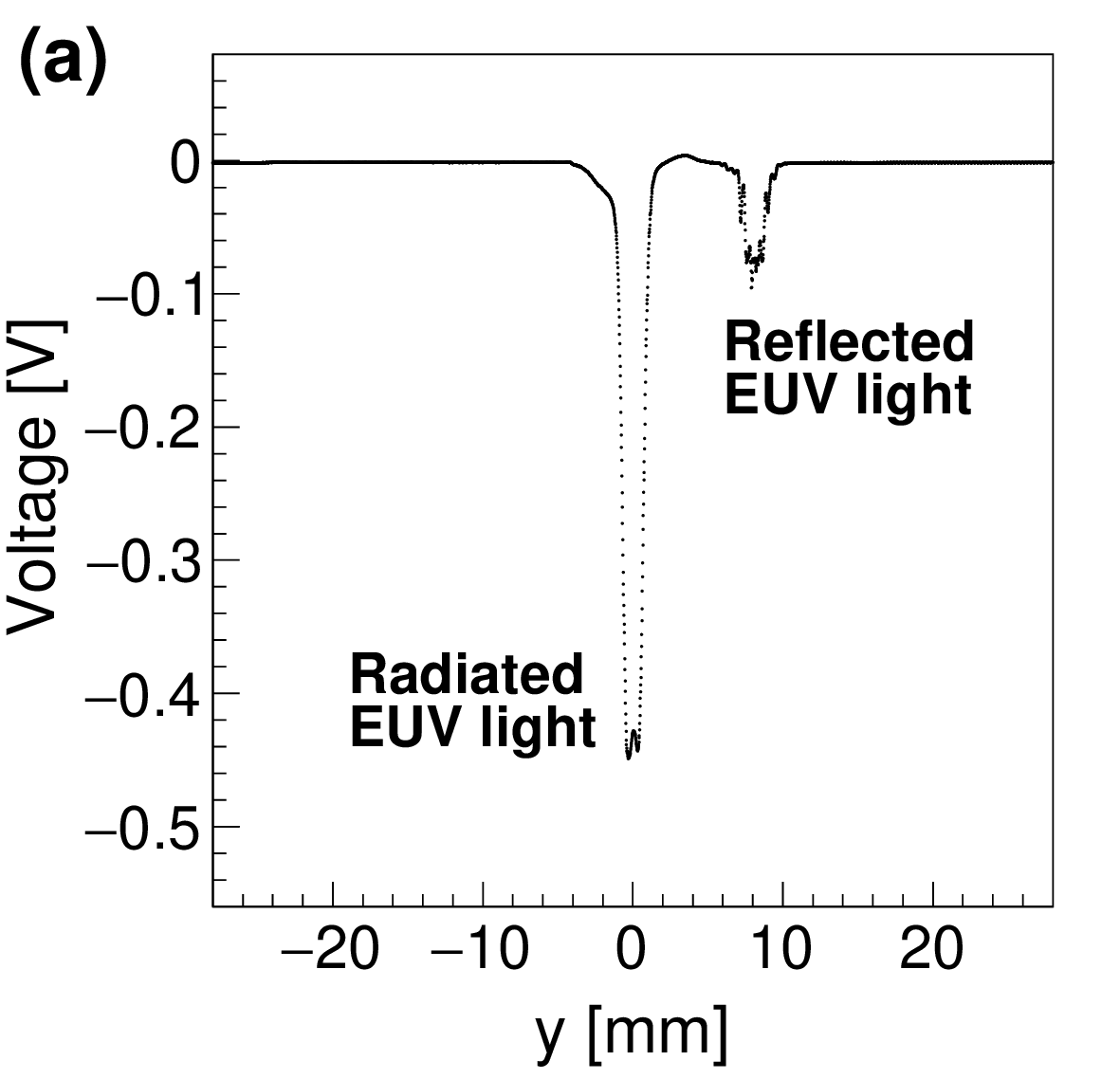}
  \end{minipage}
  \hspace{-3.0mm}
  \begin{minipage}{0.5\linewidth}
   \centering
   \includegraphics[width=6.5cm]{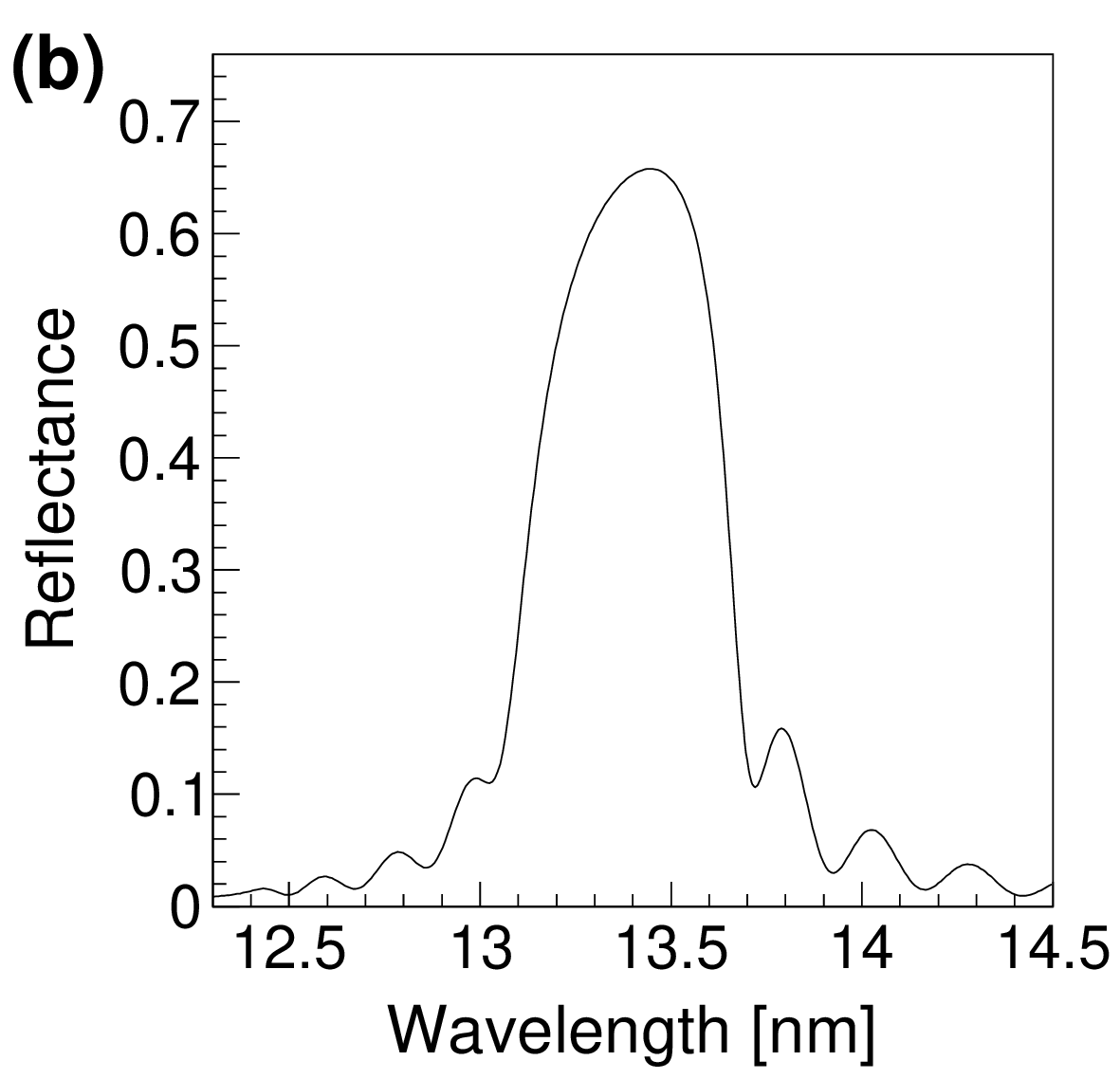}
  \end{minipage}
 \end{tabular}
 \caption{(a) Vertical profiles of radiated and reflected EUV light measured by the wire scanner. The vertical and horizontal axes show the signal voltage converted from the measured micro-current and the wire position based on a potentiometer, respectively. (b) Reflectance of the Mo/Si multilayer mirror as a function of the wavelength of EUV light incident on the center of the refractive surface.}
 \label{fig4}
\end{figure}

Figure~\ref{fig3} shows the experimental setup constructed at BL07A. Details of critical systems can be found in \ref{app2}. An X-ray monitor chamber also usable for EUV photons was installed $5.9$~$\mathrm{m}$ downstream of the switching mirror. This monitor contains a wire scanner that is driven by an air cylinder. It was used to precisely measure the positions, profiles, and relative fluxes of radiated and reflected EUV photon beams. In the wire scanner measurement, an increase of micro-current due to the photoelectric effect by EUV photon hits was detected at $0.2$~$\mathrm{mm}$-diameter tungsten wires, which were stretched vertically and horizontally. These wires were slowly moved for a scan in the directions perpendicular to each wire. Figure~\ref{fig4}(a) shows the data measured by the wire scanner during the adjustment of a reflected EUV light direction.

Inside the EUV mirror chamber, the Mo/Si multilayer mirror and its control system were set up on a motorized linear stage to be moved onto a radiated EUV photon beam axis. The mirror substrate with a $50$~$\mathrm{mm}$-square refractive surface was made of a $16$~$\mathrm{mm}$-thick silicon plate for better thermal conductivity in a vacuum. The refractive surface was cylindrically polished with a vertical curvature radius of $16.7$~$\mathrm{m}$, corresponding to the distance to the Compton scattering point, because the radiated EUV photon beam at BL07A was divergent only vertically in the propagation after the switching mirror. The RMS roughness of the polished surface reached $0.2$~$\mathrm{nm}$ per $80$~$\mathrm{\mu m}$-square area in the inspection by an atomic force microscope. The refractive surface was coated by forty periodic layers of Molybdenum and Silicon pairs. An EUV light irradiation test at another beamline confirmed its ideal performance as shown in Fig.~\ref{fig4}(b), where the maximum $R$ was $65.8$\% compared with the design value of $68$\%. The Mo/Si mirror was mounted on a water-cooled pure copper holder through a thin indium sheet. The heat load on the mirror mainly comes from the higher harmonic waves of the undulator radiation, but its amount is estimated to be at most a few $\mathrm{W}$. The mirror holder was attached to precision rotary stages that moved vertically and horizontally with the resolutions of $0.00024^\circ$ and $0.0004^\circ$, respectively.

The properties of $\gamma$-rays were examined by the detectors installed in the atmosphere behind the switching mirror. An $E_\gamma$ spectrum for the $\gamma$-rays that had passed through beamline structure materials was measured by the electromagnetic calorimeter made by assembling nine pure PbWO$_4$ (PWO) crystals into a size of $60 \times 60 \times 200$~$\mathrm{mm}$ ($22.5$ radiation lengths). Scintillation light proportional to a $\gamma$-ray hit energy was transformed to an amplified electric signal by a two-inch photomultiplier tube (Hamamatsu H7195-Y003) attached to the rear end of the PWO crystal assembly. In the studies for EUVCS, its high-voltage was suppressed to $-1400$~$\mathrm{V}$ compared with a normal operating voltage of $-2000$~$\mathrm{V}$ to cover a full $E_\gamma$ spectrum within the dynamic range of energy measurement. Thus, the $E_\gamma$ resolution in the demonstration experiment was worse than the original performance, showing $4.3$\% at the Compton edge. (See more details in \ref{app2}.) In actual use with high-rate hits, this calorimeter worked stably by adopting a booster method for the photomultiplier voltage-divider circuit. Additionally, a $3$~$\mathrm{mm}$-thick plastic scintillator was set up as a charge-veto counter most upstream of the $\gamma$-ray detector system.

\section{Results of the demonstration experiment} \label{sec3}

In the demonstration experiment to confirm the $\gamma$-ray production by EUVCS, a four-quadrant slit upstream of the X-ray monitor chamber was set to pass only a $5$~$\mathrm{mm}$-square central part of the undulator radiation, because the SRW \cite{srwsim} and EUV light propagation \cite{shadow01, shadow02} simulations suggested that the first harmonic wave of $92$~$\mathrm{eV}$ should be concentrated in a narrow region of $\sigma \approx 0.8$~$\mathrm{mm}$ at the slit both vertically and horizontally. This setting was helpful to avoid the undesired spread of an undulator radiation profile at the wire scanner and the additional heat load on the Mo/Si multilayer mirror. Then, the direction of EUV light reflected back using the Mo/Si mirror was adjusted by operating the precision rotary stages so that a profile of the reflected beam should overlap with that of the radiated beam at the wire scanner. (See Fig.~\ref{fig4}(a).) Further adjustment of the rotary stages was performed using a similar wire detector equipped upstream of the switching mirror. These two-step adjustments of the reflected EUV light positions from far to near the scattering point were essential for successfully achieving the head-on collision of EUV photons and electrons.

The energy and timing information from the $\gamma$-ray detectors was recorded using a VME-based data acquisition (DAQ) system. The whole charge of a signal pulse from the PWO calorimeter was integrated within a gate width of $300$~$\mathrm{ns}$ using a $12$-bit ADC board (CAEN V792). After optimizing the reflected EUV light direction, $8$ million events were recorded by self-triggering with PWO calorimeter signals. In addition, another $8$ million events were collected as a background sample by preventing the EUV light reflection process at an optical shutter. For further analysis, events having no coincident hit at the charge-veto counter were selected based on the ADC information in both samples.

\begin{figure}[b]
   \centering
   \includegraphics[width=8.1cm]{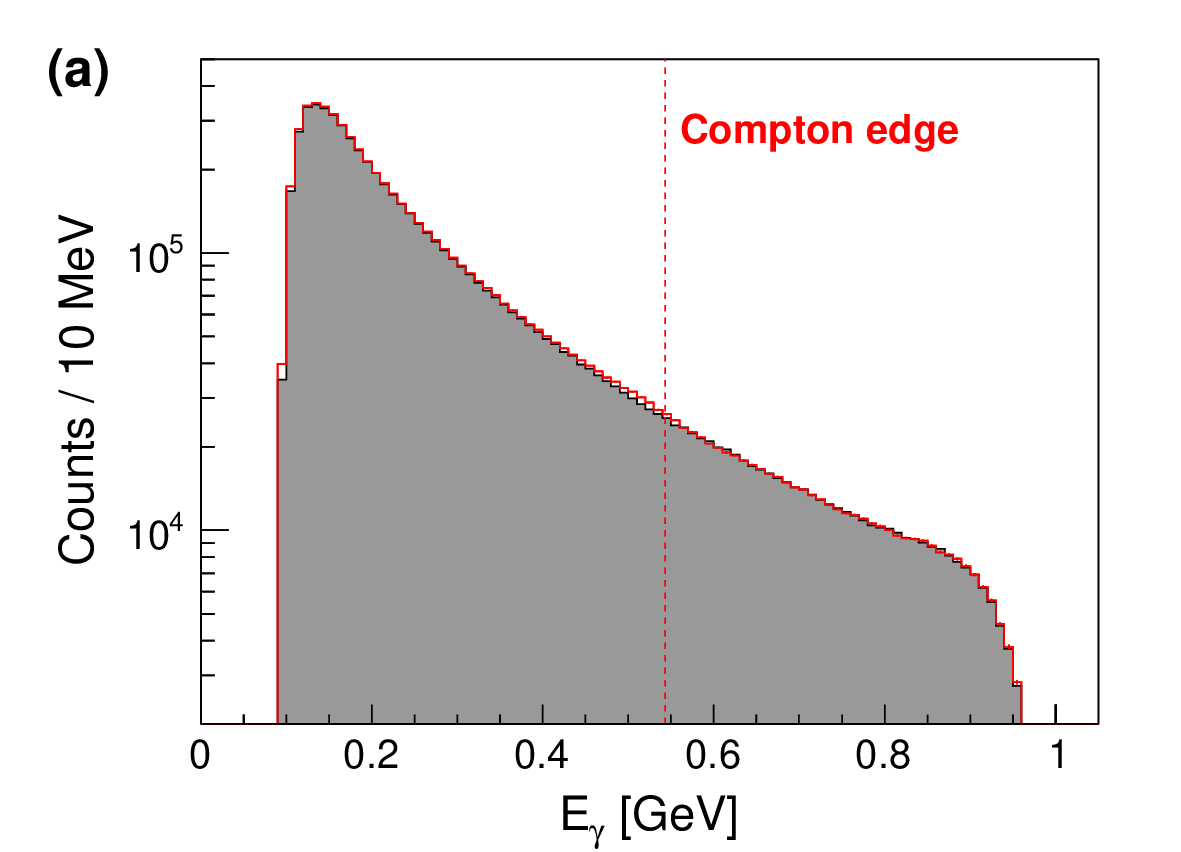}
   \centering
   \includegraphics[width=8.1cm]{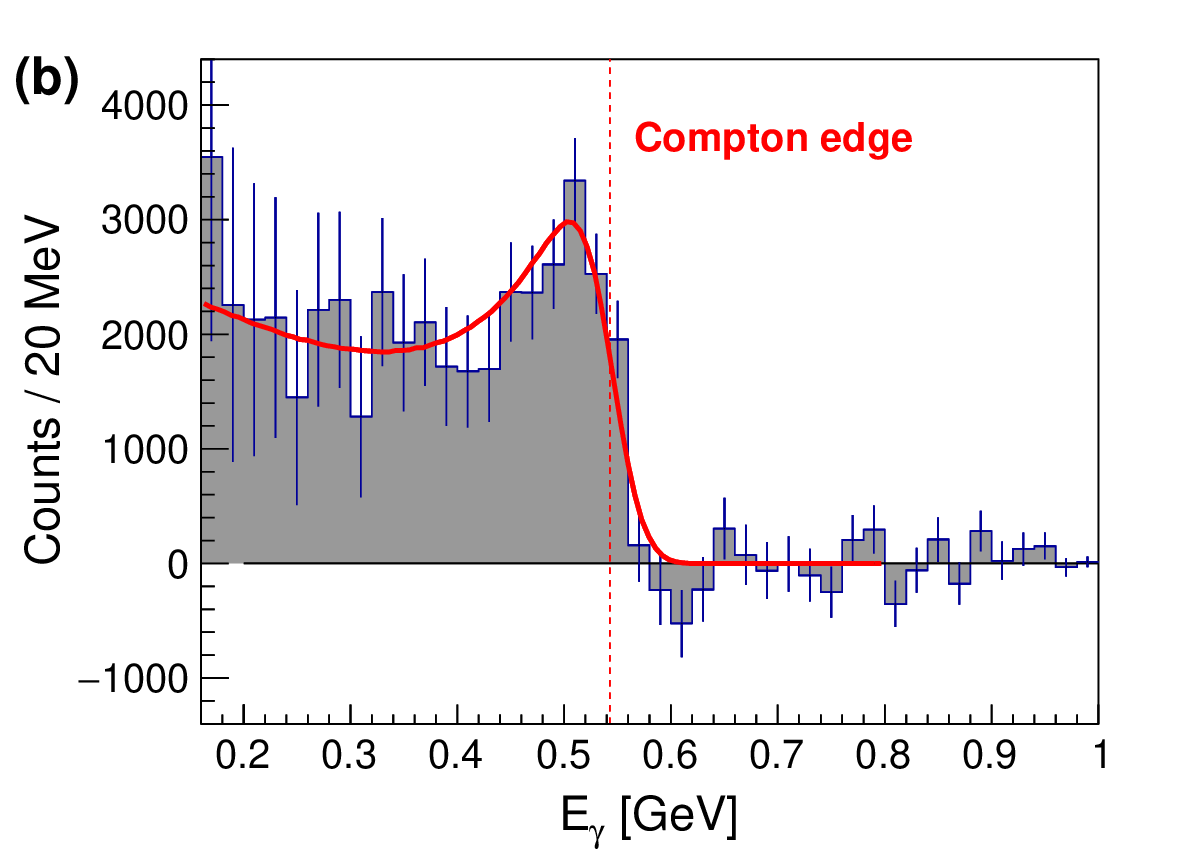}
 \caption{(a) Red open and gray filled (scaled) histograms show the $E_\gamma$ spectra measured in the demonstration experiment with and without EUV light reflection, respectively. (b) A gray filled histogram shows the difference of two spectra in (a). A red thick curve is a fit result of the theoretically calculated spectrum smeared by the calorimeter energy resolution.}
 \label{fig5}
\end{figure}

Figure~\ref{fig5}(a) shows the $E_\gamma$ spectra measured with and without the EUV light reflection. As recognized from the exponentially decreasing distribution in the case without the EUV light reflection (gray filled histogram), the $\gamma$-rays caused by bremsstrahlung radiation from the high-energy electrons passing through the residual gas inside the storage ring are inevitably present as a large background. Since the bremsstrahlung radiation should emit $\gamma$-rays up to the ring energy ($0.949$~$\mathrm{GeV}$), the $E_\gamma$ calibration was done by fitting a function created by multiplying exponential and complementary error functions to the highest edge of each ADC distribution. The lack of events in the lowest energy region around $0.108$~$\mathrm{GeV}$ or less is due to the discriminator threshold for the self-triggering signal of the PWO calorimeter. When this bremsstrahlung spectrum was overlaid on the $E_\gamma$ spectrum measured with the EUV light reflection (red open histogram with statistical error bars), the former spectrum was scaled so that the event count in $0.605$--$1.013$~$\mathrm{GeV}$ should be normalized to the corresponding count of the latter spectrum, which was expected to have no contribution from EUVCS at energies above the predicted Compton edge. As indicated in Fig.~\ref{fig5}(a), a small excess over the bremsstrahlung background was observed below the Compton edge in the $E_\gamma$ spectrum with the EUV light reflection.

Figure~\ref{fig5}(b) shows the $E_\gamma$ spectrum of the observed excess after subtracting the gray filled histogram (scaled) from the red open histogram in Fig.~\ref{fig5}(a). The bin size is changed to every $20$~$\mathrm{MeV}$ due to limited statistics. Statistical uncertainties are appropriately assigned to each bin by taking into account the bin entry counts of the two original spectra and the normalization uncertainty. The region below $0.160$~$\mathrm{GeV}$ is not plotted because the original $E_\gamma$ distributions show a rapid spectral change around the discriminator threshold, possibly causing large systematic uncertainties in the subtracted spectrum. In addition, a simple simulation to examine the influence of electromagnetic interactions at the beamline structure materials suggests that the removal of the lower energy region is effective to reduce the contribution from such interacted $\gamma$-rays. In Fig.~\ref{fig5}(b), a clear Compton spectrum was confirmed with the shape that was consistent with the theoretical calculation in Fig.~\ref{fig2}. The excess in $0.160$--$0.560$~$\mathrm{GeV}$ was counted as $43967 \pm 3510$ events, resulting in a statistical significance of $12.5\sigma$. If the spectrum of Fig.~\ref{fig2} with the convolution of the $E_\gamma$ resolution evaluated in \ref{app2} is fitted to the observed excess only by making a normalization parameter free, the measured data are well reproduced with a reduced $\chi^2$ of $0.688$. The energy spread by passage through the beamline structure materials is sufficiently smaller than the smeared resolution.

Note that the produced $\gamma$-ray beam size is narrow as discussed in \ref{app1}. The scattering cone angle of a $\gamma$-ray is less than $1.27$~$\mathrm{mrad}$ at $E_\gamma > 0.160$~$\mathrm{GeV}$, where the energy spectrum by EUVCS has been clearly observed. (See Fig.~\ref{fig6}(b).) This angle corresponds to a radius of $12.4$~$\mathrm{mm}$ at the PWO calorimeter located $9.7$~$\mathrm{m}$ away from the scattering point. The $\gamma$-rays due to EUVCS are well confined to small angles, so there is no acceptance loss at the calorimeter, which has a transverse size of $60$~$\mathrm{mm}$ square.

$8$ million events were recorded in the data collected with the EUV light reflection, whereas the number of PWO self-triggering signals was counted as $11.8$ million events by a visual scaler. Taking the ratio of these numbers, the DAQ efficiency was estimated to be $67.6$\%. Consequently, the observed $\gamma$-ray beam flux originating from EUVCS was evaluated to be $12.8 \pm 1.0$~cps by correcting the excess count in Fig.~\ref{fig5}(b) ($43967 \pm 3510$ events) with the DAQ efficiency and using the measurement time. There are $6.0$ radiation lengths of beamline structure materials unavoidable on the $\gamma$-ray beam path, so the flux is reduced to $e^{-7/9 \times 6.0}$ or $0.94$\% at the PWO calorimeter. (See further description in \ref{app3}.) The production rate of $\gamma$-rays by EUVCS was finally obtained to be $1.4 \pm 0.1$~kcps for the energy range above $0.160$~$\mathrm{GeV}$.

The ideal $\gamma$-ray production rate by EUVCS was evaluated as follows to be compared with the experimental result. Before the injection into the original ring, EUV light is reflected twice at the switching mirror ($R=88$\%) and once at the Mo/Si multilayer mirror ($R=65.8$\%). The reflection bandwidth of the Mo/Si mirror is approximately $3.5$~$\mathrm{eV}$ based on the FWHM in Fig.~\ref{fig4}(b). Thus, the flux of injected $92$~$\mathrm{eV}$ photons reaches $7.98 \times 10^{15}$~$\mathrm{s}^{-1}$ for the undulator radiation of $4.11 \times 10^{14}$~$\mathrm{s}^{-1} \cdot (0.1\% \, \rm{bandwidth})^{-1}$. On the other hand, the electron flux was calculated as $2.18 \times 10^{18}$~$\mathrm{s}^{-1}$ by dividing the beam current ($350$~$\mathrm{mA}$) by the elementary charge. Electrons and EUV photons are bunched every $2$~$\mathrm{ns}$, so that their fluxes must be divided by $0.5 \times 10^9$~$\mathrm{s}^{-1}$ to evaluate the number of particles per bunch, defined as $N_e$ and $N_{EUV}$, respectively.

The EUV light propagation until the injection was simulated using the ray-tracing software SHADOW \cite{shadow01, shadow02} to confirm the validity of the refractive surface design for the Mo/Si multilayer mirror. Since the first-harmonic undulator radiation of $92$~$\mathrm{eV}$ photons has only small divergences of $\sigma=0.04$ and $0.08$~$\mathrm{mrad}$ in the vertical and horizontal directions, respectively, the transverse spread of the EUV photon beam returning to the storage ring does not vary over the $6$~$\mathrm{m}$-long straight section where backward Compton scattering occurs. In addition, the vertical and horizontal sizes of the EUV photon beam in this straight section are similar to those of the electron beam, which has a round shape with the RMS radius of $\approx 0.6$~$\mathrm{mm}$ because of a vertical beam shaker. Oppositely traveling electron and EUV photon bunches collide with each other every $1$~$\mathrm{ns}$ at $10$ places along the straight section, namely with a rate $f_{coll}$ of $10^{10}$~$\mathrm{s}^{-1}$.

The luminosity $L$ for the electron and EUV photon bunch collision is expressed by
\begin{equation} \label{eq2}
  L = \frac{f_{coll} N_e N_{EUV}}{4 \pi \sigma_x \sigma_y}
\end{equation}
when the horizontal and vertical beam sizes ($\sigma_x$ and $\sigma_y$, respectively) are equal between the colliding bunches \cite{luminosity}. In the demonstration experiment, $\sigma_x = \sigma_y = 0.6$~$\mathrm{mm}$ gives $L = 1.54 \times 10^{32}$~$\mathrm{m}^{-2} \mathrm{s}^{-1}$ if $N_e$ and $N_{EUV}$ are averaged over all the bunches. In reality, $198$ electron bunches are filled along the ring circumference by varying the charge amount per bunch with the ratio of $3:1:3:1$ for successive $70$, $29$, $70$, and $29$ bunches. Accordingly, a correction factor of $0.941$ should be multiplied to the above luminosity, resulting in $L = 1.45 \times 10^{32}$~$\mathrm{m}^{-2}\mathrm{s}^{-1}$.

The ideal $\gamma$-ray production rate by Compton scattering of $92$~$\mathrm{eV}$ photons from $0.949$~$\mathrm{GeV}$ electrons was calculated by multiplying the luminosity and the total cross section of $330$~$\mathrm{mb}$, corresponding to the integral of Fig.~\ref{fig2}. The value to be compared with the experimental result ($1.4 \pm 0.1$~kcps) was finally evaluated as $3.2$~kcps by considering the fraction above $0.160$~$\mathrm{GeV}$ in the $E_\gamma$ spectrum of EUVCS ($67.2$\%). Their difference by a factor of $2.3$ is not large, suggesting the present experiment is quantitatively controlled. The estimated difference can arise from the misalignment between the reflected EUV photon and electron beam axes, the deterioration of the EUV light focus at the scattering point, and the decrease of the switching mirror reflectance with long use.

\section{Prospect of the developed procedure} \label{sec4}

The developed procedure of EUVCS is widely applicable at electron storage rings, which are increasing worldwide to meet the growing demand for synchrotron radiation. Currently, the highest $E_\gamma^{max}$ by LCS is obtained at SPring-8 ($E_e = 7.975$~$\mathrm{GeV}$), which is located in the same campus as NewSUBARU. In SPring-8, $E_\gamma^{max}$ has reached $2.89$~$\mathrm{GeV}$ by using $4.66$~$\mathrm{eV}$ deep UV laser light \cite{gevgex07, gevgex08}. However, SPring-8 has a near-future plan to decrease the ring energy to $6$~$\mathrm{GeV}$ (SPring-8-II \cite{spring82}), resulting in the significant drop of $E_\gamma^{max}$ with LCS. Here, if our procedure of EUVCS is adopted in SPring-8-II, $E_\gamma^{max}$ will be raised up to $5.37$~$\mathrm{GeV}$ by injecting $92$~$\mathrm{eV}$ EUV light. This is an attractive opportunity for hadron photoproduction experiments such as the study of large-mass exotic hadrons. (See more description about the obtainable energy spectrum in \ref{app1}). Additionally, the new storage ring NanoTerasu ($E_e = 2.998$~$\mathrm{GeV}$) \cite{nanoterasu} is another candidate to conduct the $\gamma$-ray beam production using EUVCS. The emission of $92$~$\mathrm{eV}$ EUV light from an undulator is more suitable at the low energy rings in the range of $1$--$3$~$\mathrm{GeV}$. In the case of NanoTerasu, $E_\gamma^{max}$ reaches $2.42$~$\mathrm{GeV}$ with the injection of $92$~$\mathrm{eV}$ EUV light. This $E_\gamma^{max}$ is comparable to that of the current LCS facility in SPring-8. Even at other low energy rings, EUVCS must be usable for fundamental physics research to reveal the non-perturbative nature of quantum chromodynamics (QCD) \cite{xcsusage}. 

The remaining challenge to put EUVCS into practical use is how to increase the production rate of $\gamma$-rays. Fortunately, new storage rings like SPring-8-II and NanoTerasu provide an improved beam emittance to increase the brightness of synchrotron radiation. The available electron beam size is getting narrow, so that the luminosity of EUVCS must be raised up as recognized from Eq.~\ref{eq2}. In the case of SPring-8-II, the horizontal and vertical electron beam sizes at a $4$~$\mathrm{m}$-long straight section are designed to be $20$ and $4$~$\mathrm{\mu m}$, respectively. These values are much smaller than the corresponding size of $0.6$~$\mathrm{mm}$ in our demonstration experiment. A SRW simulation \cite{srwsim} for SPring-8-II, whose electron beam current will be $200$~$\mathrm{mA}$, also suggests that it is possible to obtain the EUV light flux similar to that in the demonstration experiment by assuming an undulator with a large K-value. The emission of a longer wavelength radiation like the EUV light compared with the targeted wavelengths at SPring-8-II is achievable by making the K-value large \cite{kvalue1, kvalue2}. Instead, an extra care for the heat load due to the higher harmonic radiation will be needed in this case. According to the simulation by SHADOW \cite{shadow01, shadow02}, reflected EUV light can also be focused to the electron beam size although the interaction with only one electron bunch is allowed around the focus. By taking into account all the factors related to the developed method, the $\gamma$-ray production rate by EUVCS may reach $10^5$--$10^6$~cps at SPring-8-II. Proposals with more solid design, calculations, and simulations for fourth-generation storage rings (e.g., SPring-8-II, NanoTerasu, HEPS \cite{heps}, Korea-4GSR \cite{k4gsr}) will be considered in separate articles.

\section{Summary} \label{sec5}

We have developed a new innovative method to produce a $\gamma$-ray beam by the backward Compton scattering of $92$~$\mathrm{eV}$ EUV photons (EUVCS). The EUV photons are obtained from an undulator in a storage ring and are reinjected into the original ring by their backward reflection using a Mo/Si multilayer mirror. In addition to the system to reflect EUV photons, we have constructed necessary detectors at the beamline BL07A of the $1$~$\mathrm{GeV}$ storage ring NewSUBARU: the monitor system of the EUV light position to precisely adjust the direction of the reflected beam on the radiated beam axis, the electromagnetic calorimeter made of PWO crystals to measure the energy spectrum of produced $\gamma$-rays, and so on. In the demonstration experiment, a clear excess of $43967 \pm 3510$ events ($12.5 \sigma$) due to EUVCS was observed over a bremsstrahlung background by the inspection of the measured $\gamma$-ray energy spectrum. The expected EUVCS spectrum with the Compton edge of $0.543$~$\mathrm{GeV}$ fits the observed excess, showing a reduced $\chi^2$ of $0.688$. The present result is significant as the first observation of a $\gamma$-ray beam produced by the backward Compton scattering of reflected EUV light at a synchrotron radiation accelerator, whose demand is increasing. The incident EUV light has the highest (and much higher) energy to date among the seed lights used in the existing Compton scattering facilities. The new $\gamma$-ray beam source established by this work paved the way to reach high energies approaching the storage-ring energy. Our achievement is unique because of getting incident EUV light from an undulator inserted into the same beamline used for Compton scattering. This method is useful for hadron photoproduction experiments at various facilities like SPring-8-II and NanoTerasu. Although the $\gamma$-ray beam flux due to EUVCS in the demonstration experiment has been estimated to be $1.4 \pm 0.1$~kcps in the energy region above $0.160$~$\mathrm{GeV}$, this beam flux must be increased a few orders of magnitude for practical applications. Since the production rate of $\gamma$-rays by EUVCS has quantitatively been understood from the performance of each optical component, further development towards higher fluxes should be reasonably possible, for instance, by increasing the luminosity of electron and EUV photon bunch collisions with reduced transverse beam sizes in the new storage rings.

\section*{Acknowledgment}

The experiments were performed at the beamline BL07A of NewSUBARU with the approval of the Laboratory of Advanced Science and Technology for Industry (LASTI), University of Hyogo. The authors thank H.~Hirayama, Y.~Minagawa, and other staff members at NewSUBARU for providing excellent experimental conditions with many technical supports and stable ring operations. The authors are grateful to R.~Hajima for fruitful discussions on the practical use of the developed method. This research was supported by JSPS KAKENHI Grants No.~24241035, No.~24654056, No.~18H05325 / 20K20344, No.~22H01225 / 23K22496, No.~22K18707.

\appendix

\section{Kinematics} \label{app1}

\begin{figure*}[htbp]
 \begin{tabular}{ccc}
  \begin{minipage}{0.33\textwidth}
   \centering
   \includegraphics[width=6cm]{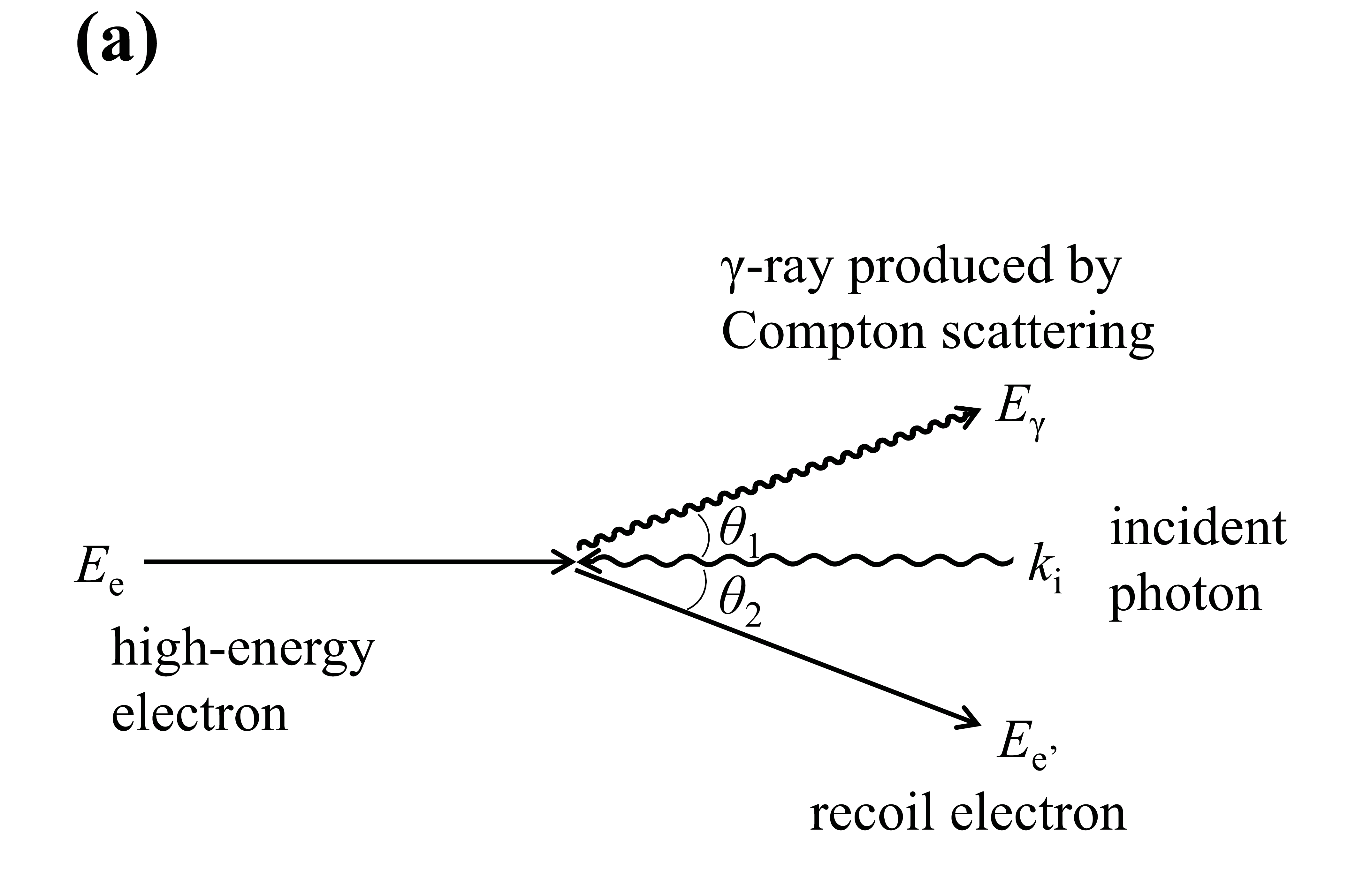}
  \end{minipage}
  \hspace{0mm}
  \begin{minipage}{0.33\textwidth}
   \centering
   \includegraphics[width=5.3cm]{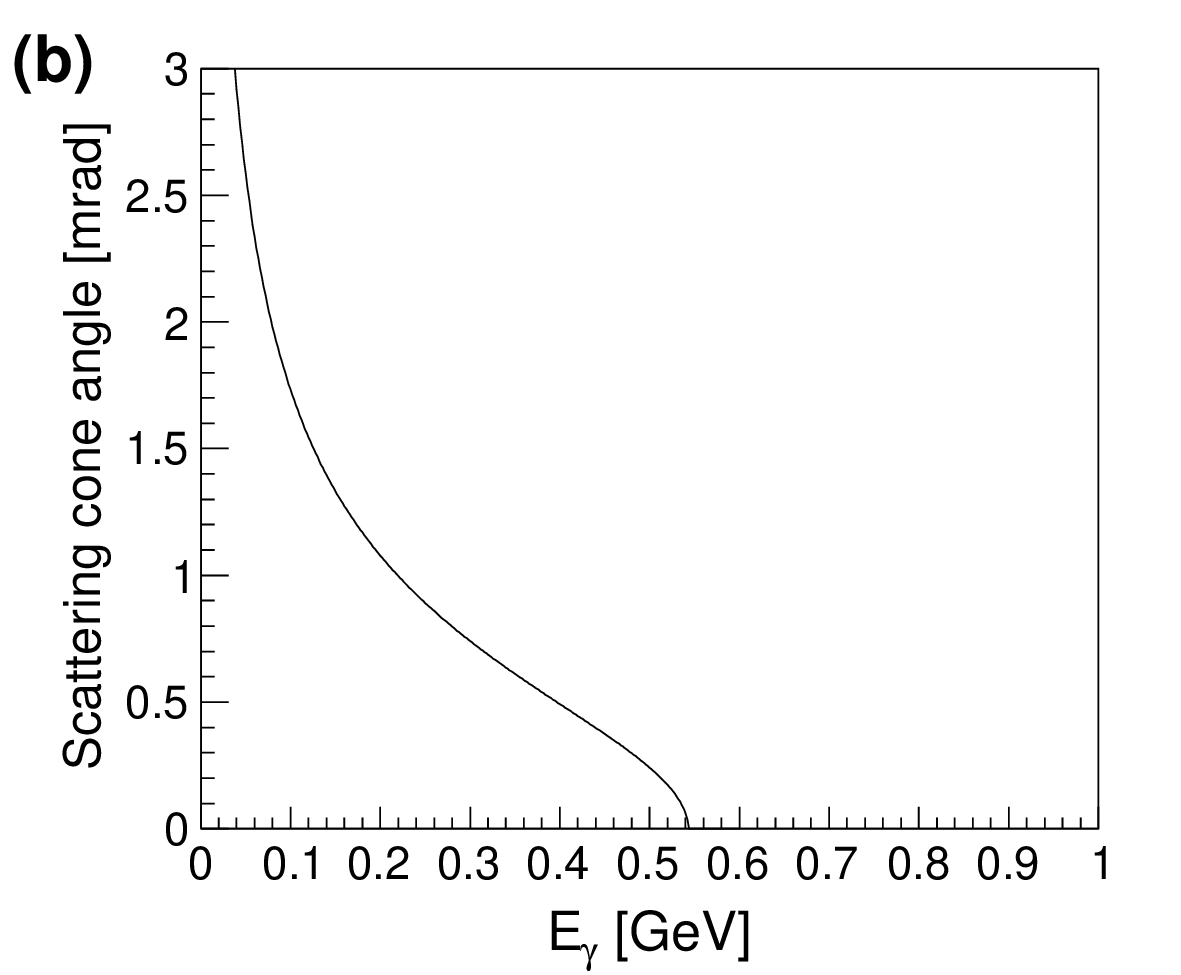}
  \end{minipage}
  \hspace{0mm}
  \begin{minipage}{0.33\textwidth}
   \centering
   \includegraphics[width=5.3cm]{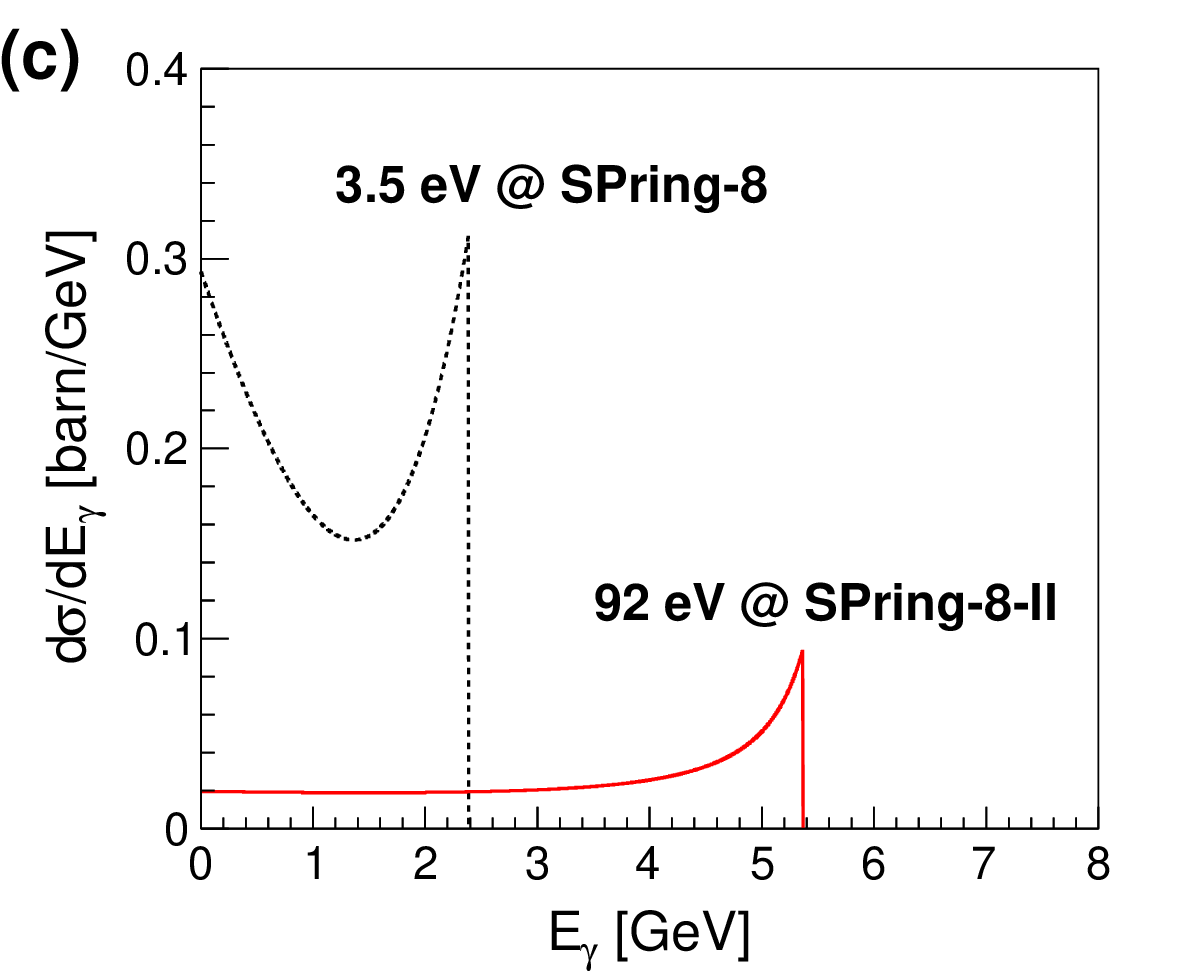}
  \end{minipage}
 \end{tabular}
 \caption{(a) A schematic diagram showing the kinematics of EUV light Compton scattering from high-energy electrons in the laboratory frame. (b) Scattering cone angles ($\theta_1$) as a function of $E_\gamma$ in the demonstration experiment of EUV light Compton scattering. (c) $E_\gamma$ spectra (differential cross sections) for EUV light Compton scattering at SPring-8-II (red solid line) and laser Compton scattering at SPring-8 (black dashed line).}
 \label{fig6}
\end{figure*}

The kinematics of Compton scattering off high-energy electrons are derived from the four-momentum conservation in the laboratory frame indicated in Fig~\ref{fig6}(a). Produced $\gamma$-rays have the maximum energy $E_\gamma^{max}$, given in Eq.~\ref{eq1}, at $\theta_1 = \theta_2 = 0^\circ$. If $\theta_1$ and $\theta_2$ are not zero, the four-momentum conservation results in
\begin{equation} \label{eq3}
  \cos \theta_1 = \frac{E_\gamma (E_e + k_i) - k_i (E_e + P_e)}{E_\gamma (P_e - k_i)} ,
\end{equation}
where $P_e$ is the initial electron momentum, represented by $\sqrt{E_e^2 - m_e^2}$. Figure~\ref{fig6}(b) shows this relation for the demonstration experiment.

Differential cross sections for Compton scattering of low-energy photons from high-energy electrons are calculated by the Klein-Nishina formula or the leading-order diagrams in quantum electrodynamics (QED) \cite{lcspre01, lcspre02, lcskin01, lcskin02, landau}. Based on such a calculation, the $E_\gamma$ spectrum for the Compton scattering of $92$~$\mathrm{eV}$ EUV light (EUVCS) at the $6$~$\mathrm{GeV}$ electron storage ring, SPring-8-II, is plotted as shown by the red solid line of Fig~\ref{fig6}(c). In contrast, the black dashed line of the same figure represents the corresponding spectrum for the usual operating condition of LCS at the SPring-8 LEPS and LEPS2 beamlines \cite{gevgex07, gevgex08}, where $3.49$~$\mathrm{eV}$ ultraviolet laser light ($\lambda = 355$~$\mathrm{nm}$) is scattered by $8$~$\mathrm{GeV}$ electrons. EUVCS in SPring-8-II provides much higher energy $\gamma$-rays than LCS in SPring-8. In Fig~\ref{fig6}(c), the total cross section for EUVCS is smaller than that for LCS by a factor of $3.4$.

\section{Developed systems} \label{app2}

\begin{figure}[htbp]
 \centering
 \includegraphics[width=12cm]{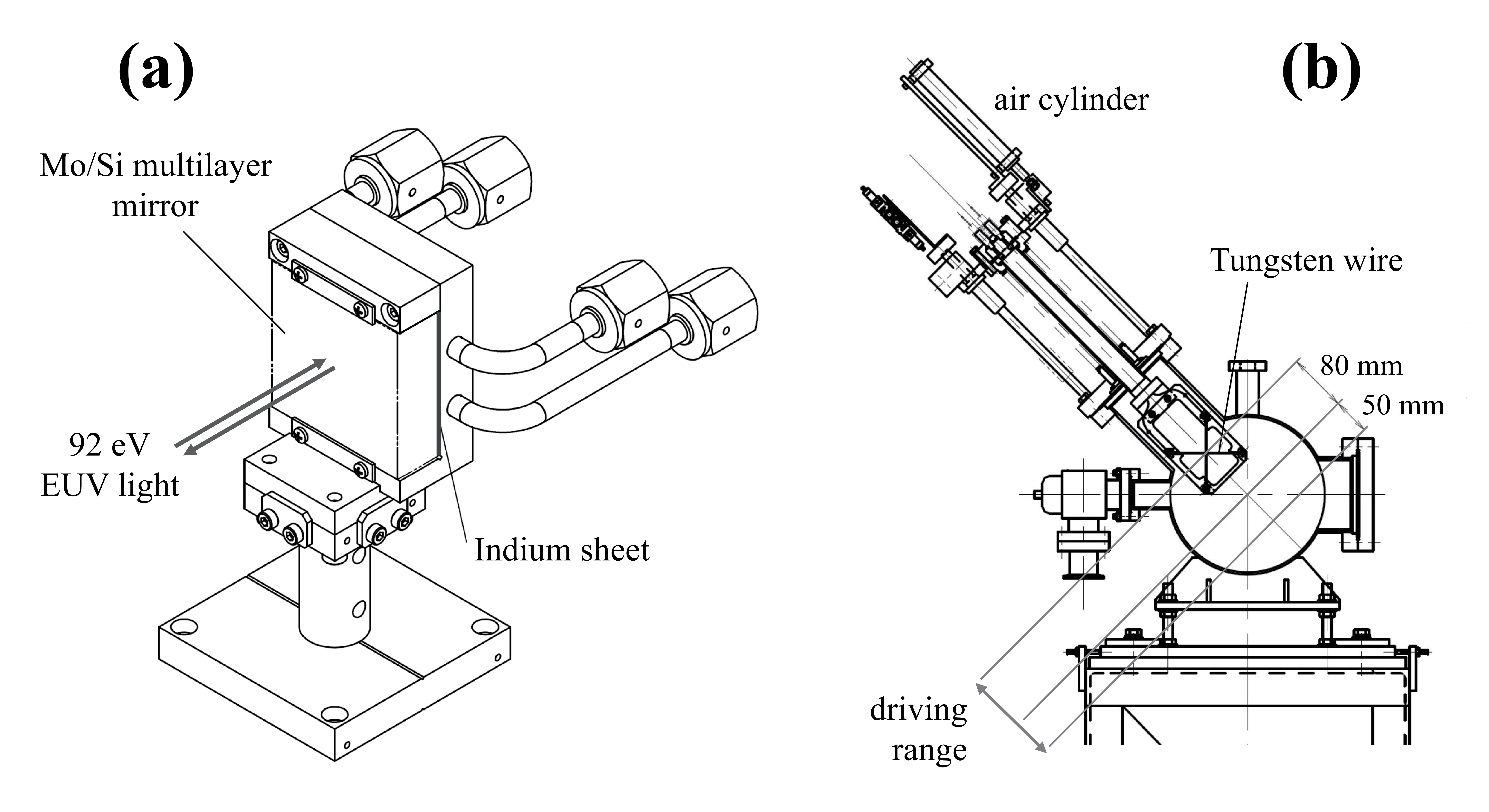}
 \caption{Systems developed for the $\gamma$-ray production by EUVCS. (a) A three-dimensional view of the Mo/Si multilayer mirror and its holder. (b) A cross-sectional view of the wire scanner mounted in the X-ray monitor chamber. This figure is a view from downstream.}
 \label{fig7}
\end{figure}

A silicon substrate of the Mo/Si multilayer mirror was produced by Crystal Optics Inc. The accuracy of its refractive surface shape is excellent, showing the peak-to-valley difference relative to the cylindrical design to be within $134$~$\mathrm{nm}$. This surface was finally polished by magnetorheological finishing. Then, forty equal-periodic layers of Molybdenum and Silicon pairs were coated on the refractive surface by NTT-AT Corp.~to achieve the Bragg reflection of $92$~$\mathrm{eV}$ photons with an incident angle of $0^\circ$. The uniformity of the coated layers was good enough at $\pm0.2$\% in a $40$~$\mathrm{mm}$-square effective area, and its validity was confirmed by a reflectance test with the variation of an EUV light irradiation point on the refractive surface. The Mo/Si multilayer mirror was attached to a holder by squeezing an indium sheet between them to ensure thermal conduction. The holder, shown in Fig.~\ref{fig7}(a), was made of pure copper with water-cooling paths that were connected to flexible metal hoses leading to a chiller outside the mirror chamber. The vertical and horizontal rotation angles of the Mo/Si mirror were controlled by two precision rotary stages (MVSA07A-RB-1F and MVRA05A-W-1F, Kohzu Precision), on which the holder was placed so as to rotate the mirror around its center. The entire mirror system is set up inside the large mirror chamber, which has an area of $820 \times 1120$~$\mathrm{mm}^2$ with a height of $610$~$\mathrm{mm}$, to provide a mirror retraction space during other experiments.

The wire scanner was constructed as shown in Fig.~\ref{fig7}(b). Tungsten wires were strung diagonally across the $80$~$\mathrm{mm}$-square area of a frame moving at an angle of $45^\circ$ to the vertical in a plane perpendicular to the beam of undulator radiation. The vertical and horizontal wires measure the horizontal and vertical profiles of an EUV photon (and also X-ray) beam, respectively, with a drive of the frame for a distance of $130$~$\mathrm{mm}$. Radiated and reflected EUV photons can be detected simultaneously by the increase of micro-current measured with a pico-ampere meter (6482/J, Keithley). The wire frame is driven by a low-speed air cylinder, which moves at a rate of $13.2$~$\mathrm{mm/s}$ with the record of a wire position using a potentiometer. Another wire detector has been installed upstream of the switching mirror as the original beamline equipment to monitor radiated X-rays. In this detector, two tungsten wires with a diameter of $0.1$~$\mathrm{mm}$ move independently in the vertical and horizontal directions using stepping motors. In the demonstration experiment, the position of a reflected EUV photon beam was precisely searched by measuring micro-current while a tungsten wire was moved step-by-step. There is also a four-quadrant slit upstream of this wire detector. It was used to reduce the spread of undulator radiation when searching for reflected EUV light by the wire detector.

\begin{figure}[htbp]
 \centering
 \includegraphics[width=9cm]{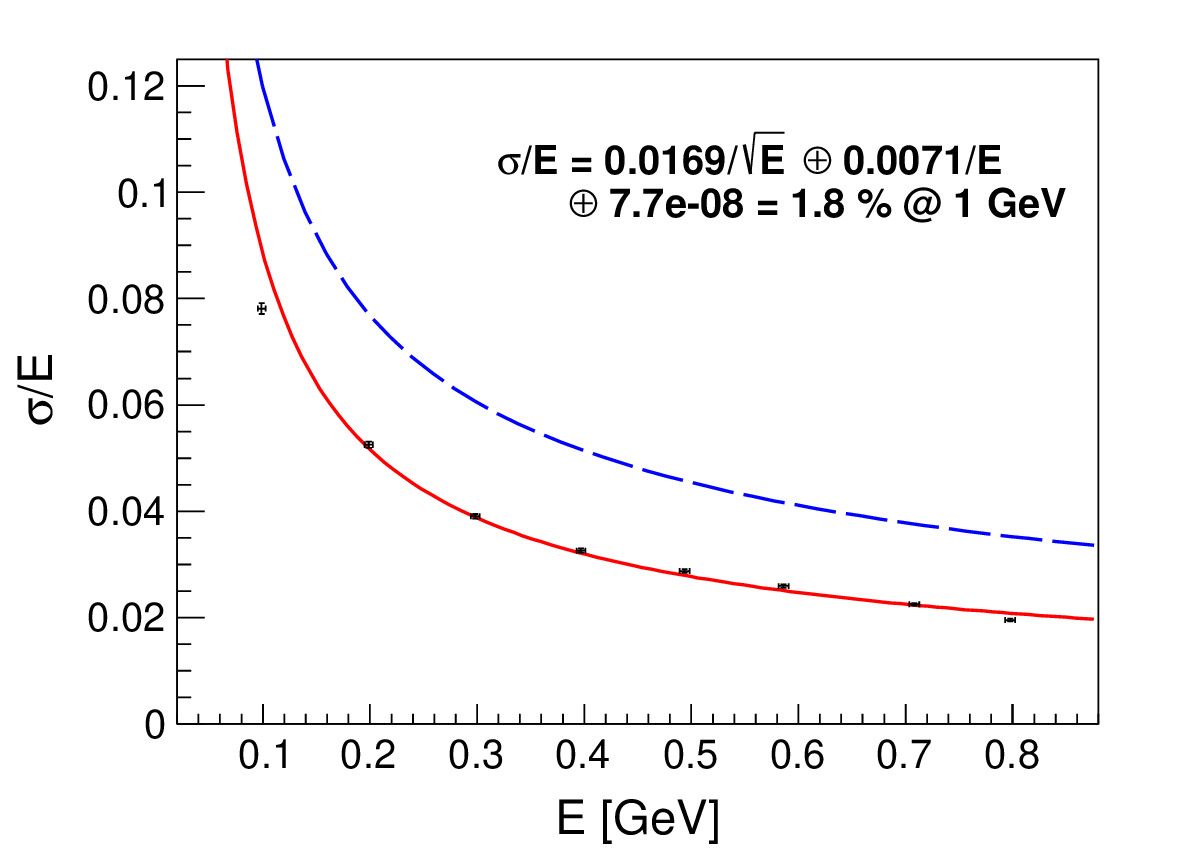}
 \caption{Closed circles with vertical bars show the energy resolutions ($\sigma / E$) and their statistical uncertainties for the PWO calorimeter operated at $-2000$~$\mathrm{V}$. Horizontal bars represent the uncertainties of the positron beam energy setting. A red solid curve and top-right texts indicate the fitting result of a resolution function. A blue dashed curve describes a modified resolution function when supplying $-1400$~$\mathrm{V}$.}
 \label{fig8}
\end{figure}

An electromagnetic calorimeter to measure the $E_\gamma$ spectrum was constructed using nine pure PWO (lead tungstate) crystals, each of which had a size of $20 \times 20 \times 200$~$\mathrm{mm}$. The assembly of them was wrapped by a reflector (ESR films) and put into a thin aluminum box for light shield. The Moli\`{e}re radius of PWO is about $20$~$\mathrm{mm}$ \cite{pwoprop}, so that a $\gamma$-ray hit entering from the $60$~$\mathrm{mm}$-square front face deposits most of the energy into the central crystal. For achieving good energy resolutions, the scintillation light outputs from nine crystals were read out together by attaching a single two-inch photomultiplier tube to the rear face of the crystal assembly. The calorimeter performance was tested using a positron beam available at Research Center for Accelerator and Radioisotope Science, Tohoku University. Closed circles in Fig.~\ref{fig8} show the energy resolutions measured at a normal operating voltage of the photomultiplier tube ($-2000$~$\mathrm{V}$). A red solid curve indicates the fitting result of a resolution function $\frac{\sigma}{E} = \frac{p_1}{\sqrt{E}} \oplus \frac{p_2}{E} \oplus p_3$ \cite{emcalgen} to them. This result indicates an energy resolution of $1.8$\% at $1$~$\mathrm{GeV}$. Because the high-voltage supplied in the demonstration experiment was lowered to $-1400$~$\mathrm{V}$, the actual energy resolutions in Fig.~\ref{fig5}(b) were worse than the red fitted curve in Fig.~\ref{fig8}. Only one resolution of $3.6$\% for the incident energy of $0.777$~$\mathrm{GeV}$ had been measured by the positron beam test with the supplied voltage of $-1400$~$\mathrm{V}$, so the fitted resolution function was modified to represent this value by updating $p_1$ in the statistical term. In Fig.~\ref{fig8}, the updated resolution function is shown by a blue dashed curve.

\section{Beamline structure materials} \label{app3}

To compare the observed $\gamma$-ray flux with the ideally expected production rate, the amount of beamline structure materials before reaching the PWO calorimeter must be estimated. Because the beamline used for the present experiments was not exclusively optimized for EUVCS, the PWO calorimeter had to be placed in the open space downstream of the switching mirror, which was set at an angle of $3^\circ$ relative to the line extended from the storage-ring straight section with the undulator. Thus, $\gamma$-rays first penetrate the switching mirror substrate made of silicon for a distance of $200$~$\mathrm{mm}$ or $2.1$ radiation lengths. Then, the $\gamma$-rays pass through the switching mirror holder made of stainless steel for $9.1$~$\mathrm{mm}$ or $0.5$ radiation lengths. Furthermore, the $\gamma$-ray path crosses a vacuum pipe for the BL07B branch at an angle of $6^\circ$ before going into the atmosphere. A length of the path through a vacuum pipe wall and flanges, all made of stainless steel, amounts to $58.9$~$\mathrm{mm}$, providing $3.4$ radiation lengths. In total, there are $6.0$ radiation lengths of materials between the Compton scattering point and the PWO calorimeter. The observed flux of $\gamma$-rays by EUVCS was corrected with the reduction factor estimated from this material amount.


%\bibliographystyle{ptephy}
%\bibliography{sample}
%
% once the .bbl file has been generated then place the text in your article.

\vspace{0.2cm}
\noindent
%For references,  note how to include DOI information from examples below. 

%This is added by T. Yoneya (editor-in-chief) on 2020/07/09.

\let\doi\relax

%without this code before the command "\begin{thebibliography}{}" , an error will be %flagged. When the bibliography is provided as separate .bib file, then this code %should be placed above the commands "\bibliographystyle{}" and "\bibliography{}" %inside the main TeX file. 

\begin{thebibliography}{9}

\bibitem{lcsfob01} O.~F.~Kulikov, Y.~Y.~Telnov, E.~I.~Filippov, and M.~N.~Yakimenko, Phys. Lett. {\bf 13}, 344--346 (1964). \\ \doi{10.1016/0031-9163(64)90040-X}

\bibitem{lcsfob02} C.~Bemporad, R.~H.~Milburn, N.~Tanaka, and M.~Fotino, Phys. Rev. {\bf 138}, B1546--1549 (1965).  \\ \doi{10.1103/PhysRev.138.B1546}

\bibitem{lcspre01} R.~H.~Milburn, Phys. Rev. Lett. {\bf 10}, 75--77 (1963). \\ \doi{10.1103/PhysRevLett.10.75}

\bibitem{lcspre02} F.~R.~Arutyunian and V.~A.~Tumanian, Phys. Lett. {\bf 4}, 176--178 (1963). \\ \doi{10.1016/0031-9163(63)90351-2}

\bibitem{mevgex01} H.~Ohgaki, {\it et al.}, IEEE Trans. Nucl. Sci. {\bf 38}, 386--392 (1991). \\ \doi{10.1109/23.289330}

\bibitem{mevgex02} D.~Nutarelli, {\it et al.}, Nucl. Instrum. Meth. A {\bf 407}, 459--463 (1998). \\ \doi{10.1016/S0168-9002(98)00068-0}

\bibitem{mevgex03} R.~Klein, {\it et al.}, Nucl. Instrum. Meth. A {\bf 486}, 545--551 (2002). \\ \doi{10.1016/S0168-9002(01)02162-3}

\bibitem{mevgex04} K.~Kawase, {\it et al.}, Nucl. Instrum. Meth. A {\bf 592}, 154--161 (2008). \\ \doi{10.1016/j.nima.2008.04.008}

\bibitem{mevgex05} S.~Amano, {\it et al.}, Nucl. Instrum. Meth. A {\bf 602}, 337--341 (2009). \\ \doi{10.1016/j.nima.2009.01.010}

\bibitem{mevgex06} H.~R.~Weller, {\it et al.}, Prog. Part. Nucl. Phys. {\bf 62}, 257--303 (2009). \\ \doi{10.1016/j.ppnp.2008.07.001}

\bibitem{mevgex07} T.~Kaneyasu, Y.~Takabayashi, Y.~Iwasaki, S.~Koda, Nucl. Instrum. Meth. A {\bf 659}, 30--35 (2011). \\ \doi{10.1016/j.nima.2011.08.047}

\bibitem{mevgex08} Y.~Taira, {\it et al.}, Rev. of Sci. Instr. {\bf 93}, 113304 (2022). \\ \doi{10.1063/5.0105238}

\bibitem{mevgex09} H.~Ohgaki, {\it et al.}, Phys. Rev. Acc. Beams {\bf 26}, 093402 (2023). \\ \doi{10.1103/PhysRevAccelBeams.26.093402}

\bibitem{mevgex10} L.X.~Liu, {\it et al.}, Nucl. Sci. Tech. {\bf 35}, 111 (2024). \\ \doi{10.1007/s41365-024-01469-3}

\bibitem{gevgex01} L.~Federici, {\it et al.}, Nuovo Cimento B {\bf 59}, 247--256 (1980). \\ \doi{10.1007/BF02721314}

\bibitem{gevgex02} D.~Babusci, {\it et al.}, Nucl. Instrum. Meth. A {\bf 305}, 19--24 (1991). \\ \doi{10.1016/0168-9002(91)90514-Q}

\bibitem{gevgex03} G.Ya.~Kezerashvili, A.M.~Milov, and B.B.~Wojtsekhowski, Nucl. Instrum. Meth. A {\bf 328}, 506--511 (1993). \\ \doi{10.1016/0168-9002(93)90667-7}

\bibitem{gevgex04} G.Ya.~Kezerashvili, A.M.~Milov, N.Yu.~Muchnoi, and A.P.~Usov, Nucl. Instrum. Meth. B {\bf 145}, 40--48 (1998). \\ \doi{10.1016/S0168-583X(98)00266-3}

\bibitem{gevgex05} G.~Blanpied, {\it et al.} (LEGS Collaboration), Phys. Rev. C {\bf 64}, 025203 (2001). \\ \doi{10.1103/PhysRevC.64.025203}

\bibitem{gevgex06} O.~Bartalini, {\it et al.}, Eur. Phys. J. A {\bf 26}, 399--419 (2005). \\ \doi{10.1140/epja/i2005-10191-2}

\bibitem{gevgex07} N.~Muramatsu, Y.~Kon, S.~ Dat\'{e}, Y.~Ohashi, {\it et al.}, Nucl. Instrum. Meth. A {\bf 737}, 184--194 (2014). \\ \doi{10.1016/j.nima.2013.11.039}

\bibitem{gevgex08} N.~Muramatsu, M.~Yosoi, T.~Yorita, Y.~Ohashi, {\it et al.}, Nucl. Instrum. Meth. A {\bf 1033}, 166677 (2022). \\ \doi{10.1016/j.nima.2022.166677}

\bibitem{phyrex01} T.~Shizuma, {\it et al.}, Phys. Rev. C {\bf 100}, 014307 (2019). \\ \doi{10.1103/PhysRevC.100.014307}

\bibitem{phyrex02} A.~Tamii, {\it et al.}, Eur. Phys. J. A {\bf 59}, 208 (2023). \\ \doi{10.1140/epja/s10050-023-01081-w}

\bibitem{phyrex03} H.~Utsunomiya, {\it et al.}, Phys. Rev. C {\bf 109}, 014617 (2024). \\ \doi{10.1103/PhysRevC.109.014617}

\bibitem{phyrex04} P.~Levi~Sandri, {\it et al.}, Eur. Phys. J. A {\bf 51}, 77 (2015). \\ \doi{10.1140/epja/i2015-15077-0}

\bibitem{phyrex05} N.~Muramatsu, {\it et al.}, Phys. Rev. C {\bf 100}, 055202 (2019). \\ \doi{10.1103/PhysRevC.100.055202}

\bibitem{phyrex06} B.~King and N.~Elkina, Phys. Rev. A {\bf 94}, 062102 (2016). \\ \doi{10.1103/PhysRevA.94.062102}

\bibitem{phyrex07} J.~K.~Koga and T.~Hayakawa, Phys. Rev. Lett. {\bf 118}, 204801 (2017). \\ \doi{10.1103/PhysRevLett.118.204801}

\bibitem{phyrex08} T.~Maruyama, T.~Hayakawa, and T.~Kajino, Sci. Rep. {\bf 9}, 51 (2019). \\ \doi{10.1038/s41598-018-37096-3}

\bibitem{iappex01} Y.X.~Yang, {\it et al.}, Rad. Phys. Chem. {\bf 218}, 111599 (2024). \\ \doi{10.1016/j.radphyschem.2024.111599}

\bibitem{iappex02} R.~Hajima, {\it et al.}, Nucl. Instrum. Meth. A {\bf 608}, S57--S61 (2009). \\ \doi{10.1016/j.nima.2009.05.063}

\bibitem{iappex03} T.~Hayakawa, {\it et al.}, J. Nucl. Sci. Technol. {\bf 53}, 2064--2071 (2016). \\ \doi{10.1080/00223131.2016.1194776}

\bibitem{lcskin01} A.~D’Angelo, {\it et al.}, Nucl. Instrum. Meth. A {\bf 455}, 1--6 (2000). \\ \doi{10.1016/S0168-9002(00)00684-7}

\bibitem{lcskin02} C.~Sun, and Y.~K.~Wu, Phys. Rev. ST Acc. Beams {\bf 14}, 044701 (2011). \\ \doi{10.1103/PhysRevSTAB.14.044701}

\bibitem{xcspre01} V.~Nelyubin, M.~Fujiwara, T.~Nakano, and B.~Wojtsekhowski, Nucl. Instrum. Meth. A {\bf 425}, 65--74 (1999). \\ \doi{10.1016/S0168-9002(98)01396-5}

\bibitem{xcspre02} R.~Hajima, and M.~Fujiwara, Phys. Rev. Acc. Beams {\bf 19}, 020702 (2016). \\ \doi{10.1103/PhysRevAccelBeams.19.020702}

\bibitem{xcspre03} H.~Ohkuma, {\it et al.}, Proc.~of IPAC2014, 941--943 (2014). \\ \doi{10.18429/JACoW-IPAC2014-TUOCA03}

\bibitem{higs}     S.~F.~Mikhailov, {\it et al.}, Proc.~of IPAC2021, 1522--1524 (2021). \\ \doi{10.18429/JACoW-IPAC2021-TUPAB067}

\bibitem{mlmirror} T.~Feigl, S.~Yulin, N.~Benoit, and N.~Kaiser, Microelectron. Eng. {\bf 83}, 703--706 (2006). \\ \doi{10.1016/j.mee.2005.12.033}

\bibitem{nima314-15} G.Ya.~Kezerashvili, A.P.~Lysenko, Yu.M.~Shatunov and, P.V.~Vorobyov, Nucl. Instrum. Meth. A {\bf 314}, 15--20 (1992). \\ \doi{10.1016/0168-9002(92)90494-O}

\bibitem{nsbl07} S.~Tanaka, S.~Suzuki, T.~Mishima, and K.~Kanda, J. Synchrotron Radiat. {\bf 28}, 618--623 (2021). \\ \doi{10.1107/S1600577520016781}

\bibitem{nsring} A.~Ando, {\it et al.}, J. Synchrotron Radiat. {\bf 5}, 342--344 (1998). \\ \doi{10.1107/S0909049597013150}

\bibitem{landau} V.B.~Berestetskii, E.M.~Lifshitz, and L.P.~Pitaevskii, ``Quantum Electrodynamics'', second edition, Elsevier (1982).

\bibitem{srwsim} O.~Chubar and P.~Elleaume, Proc. of the EPAC98 Conference, 1177--1179 (1998).

\bibitem{kvalue1} S.~Krinsky, M.L.~Perlman and R.E.~Watson, ``Characteristics of synchrotron radiation and of its sources'' in Handbook on Synchrotron Radiation (Edited by E.-E.~Koch), {\bf Vol. 1A}, 65--172, North-Holland Pub.~(1983).

\bibitem{kvalue2} M.~R.~Howells and B.~M.~Kincaid, LBL-34751 / UC-406 (1993).

\bibitem{shadow01} F.~Cerrina and M.~Sanchez~del~Rio, Handbook of Optics {\bf V} (3rd edition), Ch. 35 (Mc Graw Hill, 2009).

\bibitem{shadow02} M.~Sanchez~del~Rio, N.~Canestrari, F.~Jiang, and F.~Cerrina, J. Synchrotron Radiat. {\bf 18}, 708--716 (2011). \\ \doi{10.1107/S0909049511026306}

\bibitem{luminosity} W.~Herr and B.~Muratori, CERN-2006-002, 361--377 (2006). \\ \doi{10.5170/CERN-2006-002.361}

\bibitem{spring82} H.~Tanaka, {\it et al.}, J. Synchrotron Rad. {\bf 31}, 1420--1437 (2024). \\ \doi{10.1107/S1600577524008348}

\bibitem{nanoterasu} S.~Obara, {\it et al.}, Phys. Rev. Acc. Beams {\bf 28}, 020701 (2025). \\ \doi{10.1103/PhysRevAccelBeams.28.020701}

\bibitem{xcsusage} C.R.~Howell, {\it et al.}, J. Phys. G: Nucl. Part. Phys. {\bf 49}, 010502 (2022). \\ \doi{10.1088/1361-6471/ac2827}

\bibitem{heps} Y.~Jiao, {\it et al.}, Proc.~of IPAC2024, 1290--1293 (2024). \\ \doi{10.18429/JACoW-IPAC2024-TUPG31}

\bibitem{k4gsr} J.~Kim, {\it et al.}, Proc.~of IPAC2025, 789--792 (2025). \\ \doi{10.18429/JACoW-IPAC2025-MOPS089}

\bibitem{pwoprop} M.~Ippolitov, {\it et al.}, Nucl. Instrum. Meth. A {\bf 537}, 353--356 (2005). \\ \doi{10.1016/j.nima.2004.08.042}

\bibitem{emcalgen} C.~W.~Fabjan and F.~Gianotti, Rev. Mod. Phys. {\bf 75}, 1243--1286 (2003). \\ \doi{10.1103/RevModPhys.75.1243}

\end{thebibliography}

\end{document}